# NOMA Schemes for Multibeam Satellite Communications

Ana I. Perez-Neira[1], Marius Caus, Miguel Angel Vazquez, Nader Alagha[2]

**ABSTRACT**

Non-orthogonal multiple access (NOMA) schemes are being considered in 5G new radio developments and beyond. Although seminal papers demonstrated that NOMA outperforms orthogonal access in terms of capacity and user fairness, the majority of works have been devoted to the wireless terrestrial arena. Therefore, it is worth to study how NOMA can be implemented in other type of communications, as for instance the satellite ones, which are also part of the 5G infrastructure. Although, communications through a satellite present a different architecture than those in the wireless terrestrial links, NOMA can be an important asset to improve their performance. This article introduces a general overview of how NOMA can be applied to this different architecture. A novel taxonomy is presented based on different multibeam transmission schemes and guidelines that open new avenues for research in this topic are provided.

I.  INTRODUCTION

The 5$^{th}$ generation (5G) mobile system will help, among others, to develop today's mobile broadband use cases, the so-called enhanced mobile broadband (eMBB), by increasing the peak data rate, the area traffic capacity, the user experience data rate, and the spectral efficiency. For the improvement of these key performance indicators NOMA schemes have attracted a lot of attention [1-2]. From the theory, it is known that in certain communication channels, non-orthogonal superposition techniques could outperform orthogonal schemes, such as frequency division multiplex or time division multiplex. This happens in particular in those scenarios where users can be ordered in a natural way from the strongest to the weakest because there is a large imbalance among user link qualities. Then, they can share the same time and frequency slot if the transmitter assigns different powers. At the receiver side, interference cancellation or multiuser detection[3] (MUD), can be utilized to decode the superimposed signals.

In parallel to the 5G evolution, the ever growing demand for high data throughput has triggered research efforts to improve the spectral efficiency of satellite communications. Very high or high throughput satellites (V/HTS) [3] are emerging satellite solutions that highly demand such spectral efficiency improvements. They deploy multiple beams that tessellate the coverage area in small beam footprints; thus, allowing frequency reuse across the coverage area. By adopting three or four-frequency reuse schemes among beams it is practically guaranteed certain level of isolation between cochannel beams (i.e. much reduced interbeam interference). In this way, in spite of the frequency reuse, the interference, which comes from neighboring beams operating at the same frequency, is significantly lower than the desired signal. Thus, the user terminals can apply single user decoding (SUD) strategies, which entail treating the interbeam interference as noise, without significant performance degradation. With the aim of lowering the cost per b/s and increasing the spectral efficiency or the available system bandwidth, new systems foster more aggressive frequency reuse and the interbeam interference increases. To tackle this increase of the interference levels, existing satellite standards, as the DVB-S2x (annex E) consider multibeam precoding techniques in the forward link, which use the knowledge of the users' channels at the transmitter in order to precancell the interference. We note that the forward link denotes the communication link from the transmitter or gateway to each receiver or user terminal through the satellite.

---

[1] The first three authors are with the *Centre Tecnológic de Telecomunicacions de Catalunya*
[2] *European Space Agency (ESA)*
[3] By interferene cancellation or MUD we denominate all those receiver architectures that are able to decode multiple signals, either in parallel or sequentially.

To further encourage an increase of the spectral efficiency, this article considers the possibility of having non-orthogonal multiple access within the beams. Specifically, in the current satellite communications the hundreds of users that lie within one beam access the satellite by orthogonal channels in time, frequency or code, depending on the standard. In this article we explain how NOMA can be incorporated so that users can share the same time-frequency-code resource. As such, NOMA requires not only transmission strategies, but also effective signal detection techniques at the receivers, and ultimately, a new design of network resource allocation strategies, physical layer and access designs. Although there are papers that review precoding techniques in multibeam satellites, there is not already an appropriate tutorially-oriented article on how NOMA can be applied in this kind of communications. With this aim, this article introduces an overview taxonomy of NOMA for multibeam satellite and focuses on the main specificities that must be considered.

The rest of the article is organized as follows: Section II presents the basic idea behind NOMA and lays out the different NOMA strategies that arise in the satellite multibeam architecture. Next, Section III explains how NOMA can be integrated with precoding in order to notably improve spectral efficiency, whenever there is full channel state information (CSI) at transmission. After that, Section IV draws some conclusions and outlines further research directions.

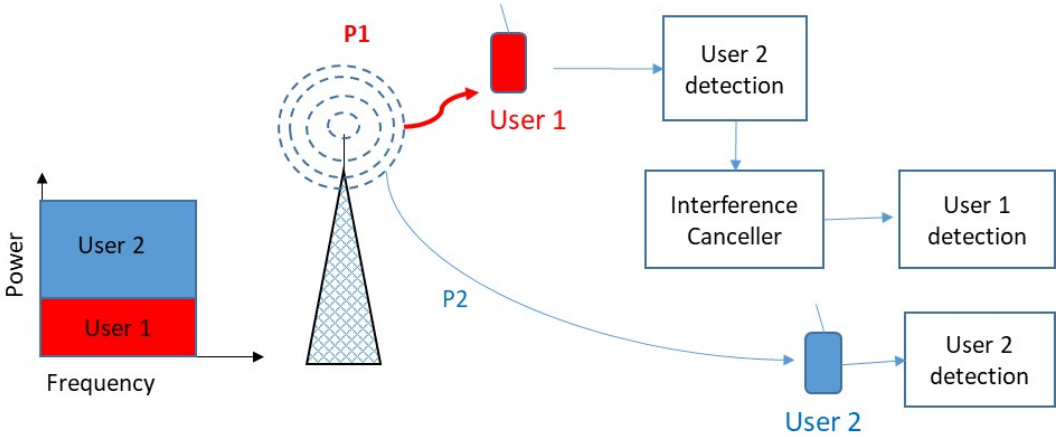

Figure 1. Downlink power-domain NOMA in a cellular scenario for the two-user case.

## II. KEY IDEAS OF THE APPLICATION OF NOMA TO MULTIBEAM SATELLITE COMMUNICATIONS

The current available NOMA techniques can be broadly divided into two categories; i.e. power-domain and code domain [1]. However, in most cases, NOMA is associated with multiplexing signals in the power domain [2]. Figure 1 sketches the main aspects of this scheme in a cellular scenario with two users. One of the users (user one in the figure) receives markedly a *stronger* signal than the other (e.g. experiences a better channel). The other user (user two in the figure) is referred to as the *weak* user, which can be, for instance, closer to the edge of the cell. In NOMA, this asymmetry is usually harnessed by assigning more transmit power to the *weak* user. As for the decoding strategy, the *weak* user recovers the corresponding message by treating the interference as noise with a SUD, while the *strong* user can recover both messages by resorting to MUD strategies. That is, the *strong* user receives also the interference strong enough so that it can be decoded first, and then performs interference cancellation or MUD to recover the desired message. We note that another possible strategy to take advantage of this asymmetric power reception is to transmit simultaneously two signals from two different layers or services (e.g. 4K high-definition services and 8K ultra-high-definition services). These two signals can be encoded with different types of error correcting code and modulation. In principle, the user who experiences higher link quality could decode both layers. This is the so-called layered division multiplexing (LDM) [4], where the signals are superimposed in the same manner as power-domain NOMA. In this article we consider also LDM strategies to be within NOMA. Note that the NOMA scheme can be applied to more than two users, whenever their channel conditions are distinct. However, in practice, the complexity of interference cancellation or MUD at the receiver increases considerably when there is more than one interference. For this reason and without loss of generality we consider in this article a two-user NOMA.

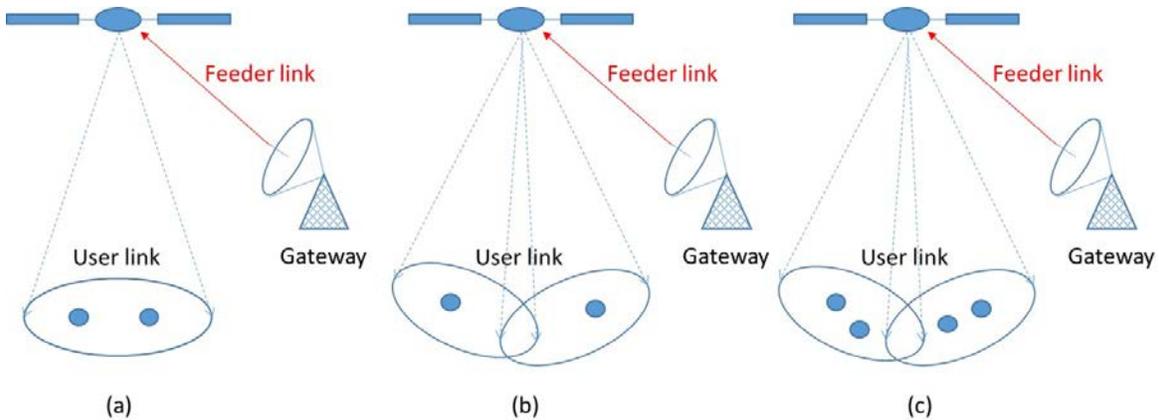

Figure 2. Different possible interfering satellite architectures for NOMA in the case of at most two simultaneous users per beam: a) isolated single beam because the rest of beams transmit with other frequencies; b) two beams sharing the frequency band and one user per beam; c) two beams sharing the frequency band and two users per beam.

The mentioned NOMA transmission scheme can be mimicked in diverse satellite multibeam schemes or architectures, which are sketched in Fig. 2. The multibeam nature of the satellite opens a range of possible multibeam NOMA strategies, which, to the authors' knowledge, is the first time that they are systematically studied within the satellite context. For the sake of clarity in the exposition, Fig. 2 isolates just one or two beams among the possible $K$ beams in a V/HTS geostationary satellite. The satellite has a transparent payload and, therefore, all the processing is done at the gateway. In this section, we will discuss the design issues for non-aggressive and aggressive frequency reuse.

## A. Non-aggressive frequency reuse among beams

First, we refer to Fig. 2a, which draws a perfect analogy between one satellite beam and one terrestrial cell. In this case, satellite communications can directly benefit from the theory developed in the wireless terrestrial NOMA framework. Figure 2a represents the situation when the frequency reuse among beams is not very aggressive (e.g. by adopting three or four frequencies) and beams do not need to cooperate, then NOMA can be applied on a per beam basis. Among the decoding strategies for NOMA, we remark that if each user knows beforehand the rate that is assigned to the interference and does not care about the errors when decoding it, a general approach is possible to achieve any rate pair. This strategy is the so-called simultaneous non-unique decoding (SND) [5], where, depending on the rate of the interference, any of the two users may treat it as noise (i.e. SUD) or may opt for decoding and regenerate it to perform interference cancellation. Figure 3 illustrates the jointly achievable rate region of two users that apply SND, which it is the union of interference as noise and simultaneous decoding (i.e. interference cancellation) regions. Figure 3a corresponds to the case where the interference is high with respect to the noise. By contrast, the rate region depicted in Fig. 3b determines the achievable rates when the interference power is much below that of the noise for user one.

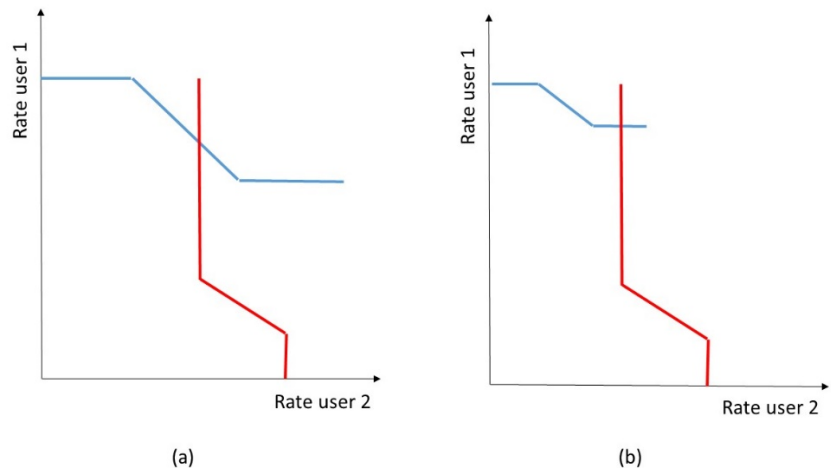

Figure 3. Rate region for the SND strategy as the intersection of the rate of 2 users: a) high interference vs. noise; b) low interference vs. noise.

Note that, within a single beam, NOMA can be envisaged as long as there is a significant signal-to-noise (SNR) imbalance among users [3]. However, in V/HTS communications, low fluctuations of the satellite channel are expected within one beam. Indeed, the crossover point of beams is usually set so that the path loss variation within the beam is confined to 3 or 4 dB. Nevertheless, the SNR imbalance comes into play due to: heterogeneous user terminals with different receiving antenna sizes and gains (e.g. on fixed and mobile terminals [6-7]), receivers of different quality, and the shadowing/blockage that arises in mobility scenarios. Essentially, the SNR imbalance appears when distinct classes of user terminal are deployed. This includes the scenario of fixed satellite services in coexistence with aeronautical or vehicular users that are connected via satellite.

In any case, note that in V/HTS, one satellite has control on a large number of beams (i.e. from hundreds to thousands in next generation VHTS) and if these beams collaborate (e.g. with precoding techniques), SNR imbalance among the satellite user terminals can appear [8]. Next, this multibeam nature is used jointly with NOMA in order to cope with an aggressive frequency reuse.

## B. Aggressive frequency reuse among beams

As we have pointed out, to unleash the full potential of multibeam satellite communication systems, it is deemed necessary to adopt two- or full-frequency reuse schemes, instead of the four-frequency reuse one. For the sake of clarity, and to better illustrate the concept of NOMA in V/HTS, first we consider in this section that two beams reuse frequency with one user per beam, as illustrated in Fig. 2b. Hence, the global multibeam system can be regarded as a set of multiple two-beam and two-user communication systems. In this case, NOMA transmission techniques shall be designed to cope with interbeam rather than the intrabeam interference. If there is no beam cooperation, we can draw an analogy with the two-user interference channel (IC), which contrasts with the broadcast channel (BC) that usually comes up in the cellular architecture. In the IC, characterized by the Han-Kobayashi (HK) capacity inner bound, close-to-optimal strategies consist in dividing each user's message into private and common message, which are sent via superposition coding. This opens the door to the so-called rate splitting approaches [9], whose implementation has to take into account that the common message is decodable by both users, while the private one has to be recovered only by its intended receiver and it is not decodable by the other one. Although user 1 (2) is not required to decode the common message intended to user 2 (1), the effect of the interference could be reduced if each user decodes its intended message as well as the common message, which conveys information from the interfering beam. Among the decoding strategies, depending on the rate of the common message and the SNR imbalance at each of the two user terminals, any of them may treat the interference as noise or may opt for jointly decoding the public messages, regenerate them, perform interference cancellation, and finally, recover the corresponding private message. In other words, the SND strategy can be used as in the non-rate splitting approach.

In order to attain practical schemes, as the general theoretical HK scheme is complex to devise, and also by taking advantage of the possible centralized control that the satellite gateway has on the beams, certain degree of cooperation between the two beams should be envisaged. If beams are allowed to cooperate in the scenario represented in Fig. 2b, then the BC model becomes now the reference. Each pair of beams can be independently processed to apply, for instance, the most suitable coding scheme depending on several aspects, such as the feedback signaling and the possibility of implementing MUD at the receive side. In [10] the authors obtain good results in front of orthogonal access in a two-frequency reuse scheme just by only exploiting partial CSI at transmission in the form of channel amplitude or SNR. In this case, interference is not completely canceled at the transmit side and thus, user terminals must jointly decode two non-orthogonal signals by resorting to interference cancellation techniques.

Remarkably, if the beams cooperate, channel imbalance between the two users results not to be so imperative. As an example, Fig. 4 illustrates that rate splitting combined with beam cooperation of [10] achieves a larger rate region than that of orthogonal schemes. In this example, it has been considered that the SNR and interference-to-noise ratio (INR) are the same for each user at each beam.

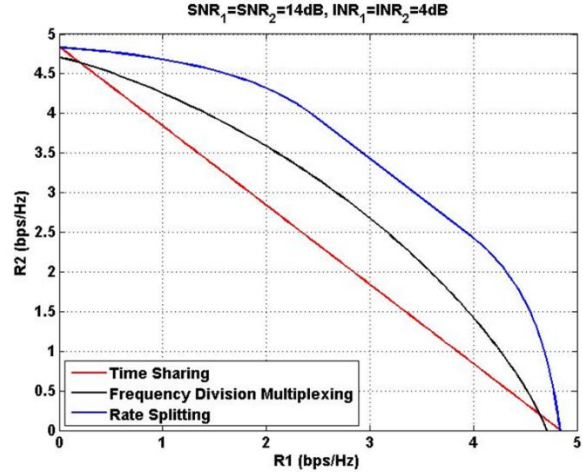

Figure 4. Achievable rate regions for different strategies that are applied to IC with two users.

In general, the better the cooperation, the better the performance is, and also challenging scenarios as the overloaded one in Fig. 2c, can be addressed. Figure 2c considers that there is full CSI at transmission, and that this allows full beam cooperation. In other words, the inputs to the multiple antennas or feeds that are on the satellite to create the spot beams are jointly processed to cope with the interference at the user terminals. We classify this new setup within the multiple-input-single-output (MISO) broadcast channel (BC) framework, assuming that the user terminals are equipped with a single antenna. Next section further expands how the precoding techniques that are appropriate for the MISO BC channel in V/HTS can be jointly designed with NOMA schemes. Indeed, the high performance receivers with MUD capabilities may increase the data rate of precoding techniques in multibeam satellite systems.

### III. NOMA AND PRECODING

Recent advances in multibeam satellite systems have identified spatial precoding as a key enabler for increasing the spatial multiplexing gain [11]. If the V/HTS controls $K$ beams, these are created with $N$ ($N \geq K$) feeds or antenna elements onboard. Spatial precoding allows the transmission over an aggressive frequency reuse system by jointly processing at the gateway the input signal to some of the beams or, alternatively to some of the feeds, in order to mitigate the interference. In this context, the user terminal can benefit from a large available bandwidth preserving a sufficiently large signal to interference and noise ratio (SINR) value. This is done with full CSI at the transmitter (i.e. channel phase and channel amplitude) and via the precoding operation and user scheduling methods.

Theoretically, the capacity achieving strategy in the MISO BC is the dirty paper coding. However, the complexity renders dirty paper coding as impractical. Alternatively, linear precoding stands as a promising solution that strikes a good balance between complexity and performance. If the gateway is aware of the channel vector of each user, then it is possible to exploit this information to precancel the interference at the transmit side, allowing the use of SUD at the receiver. Therefore, the complexity is placed at the transmitter. It is worth mentioning that linear precoding relies on the full CSI that is regularly fed back by user terminals in the return link.

Precoding is necessary in case of full-frequency reuse, where there can be up to three or four strong interfering signals. Since the complexity of MUD grows exponentially with the number of signals to be decoded, it becomes evident that precoding techniques with full CSI have to be incorporated in order to cope with the interference in aggressive frequency reuse schemes. The receiver can be just a SUD or it can be more sophisticated, which depends on the targeted scenario. In [12] the authors evaluate the achievable rates of the joint combination of precoding and MUD in the forward link of multibeam satellite system with full-frequency reuse, and when one user is served at each beam. The results show the benefits of this joint scheme that eventually can increase the current precoding performance a 100 percent. It remains an open topic to study how different joint precoding and MUD schemes compare in terms of complexity, amount of required feedback channel, robustness to imperfect channel information, and performance in different situations (i.e. unicast, multicast and broadcast).

As commented, in order to further increase the spectral efficiency, the system can be overloaded by serving more users than antenna elements in the satellite. This is the case in Fig. 2c, where two users per beam are served. This scenario is addressed in [13], where two main aspects are introduced: first, NOMA is incorporated in order to deal with the intrabeam interference (SUD and SND strategies are considered), and precoding deals with the interbeam interference; second, suitable and new scheduling schemes are studied. Concerning the scheduling, in this case, the scheduling algorithms conceived for SUD techniques are not appropriate. In this sense, new algorithms to pair users with beams are proposed. Numerical results reveal that NOMA (i.e. with SND) outperforms the conventional NOMA approach, in terms of sum rate and fairness. In addition, when the system is overloaded, the best strategy is to pair users with almost collinear channels and the lowest channel gain difference. This is precisely opposed to the best scheduling strategy for the conventional NOMA. In [13] it is also shown that the sum rate increases in spite of the user overload. In any case, the performance of a given scheduling mechanism is coupled with the elected precoding strategy, and this is a topic that deserves more research and attention.

Remarkably, more than two users per beam can be simultaneously coped with if precoding for multicast is introduced. In fact, the recent developments in framing that have appeared in the digital video and broadband S2X standard (DVB-S2X annex E) allow the implementation of the so-called multicast precoding. In multicast precoding the same information is sent to a group of let us say $N_u$ users. In this way, longer channel codes, which encompass several users, can be used to counteract the high path loss of the satellite channel. With this aim, the gateway shall design a *K* x *N* matrix (where *K* is the number of beams and *N* the number of feeds or antenna elements, $N \geq K$) in order to serve a number of $N_U$ users per beam. If, in addition, the users can exploit MUD capabilities, different NOMA alternatives can be studied. For instance, the system designer can opt for superimposing two different messages within each beam, then the precoding matrix is *K* x *N*. Another alternative is to split the rate into a single broadcasting or public transmission and the different messages that are intended for each beam. For that case, the gateway shall compute a precoding matrix of (*K +1*) x *N* elements.

The overall functioning is depicted in Fig. 5, where an example of *K* = 2 and $N_U$ = 2 has been identified as a benchmark. In a system with LDM, the aim is to send information via the simultaneous transmission of two frames superposed in the same feed element. On the contrary, in a joint broadcasting and multicasting scheme, the aim is to transmit a single frame over the coverage area together with each dedicated beam frame (i.e. a common frame is included in all feed elements). In [14] the authors show the benefits of broadcast/multicast rate splitting with precoding. The results improve those of only multicast precoding.

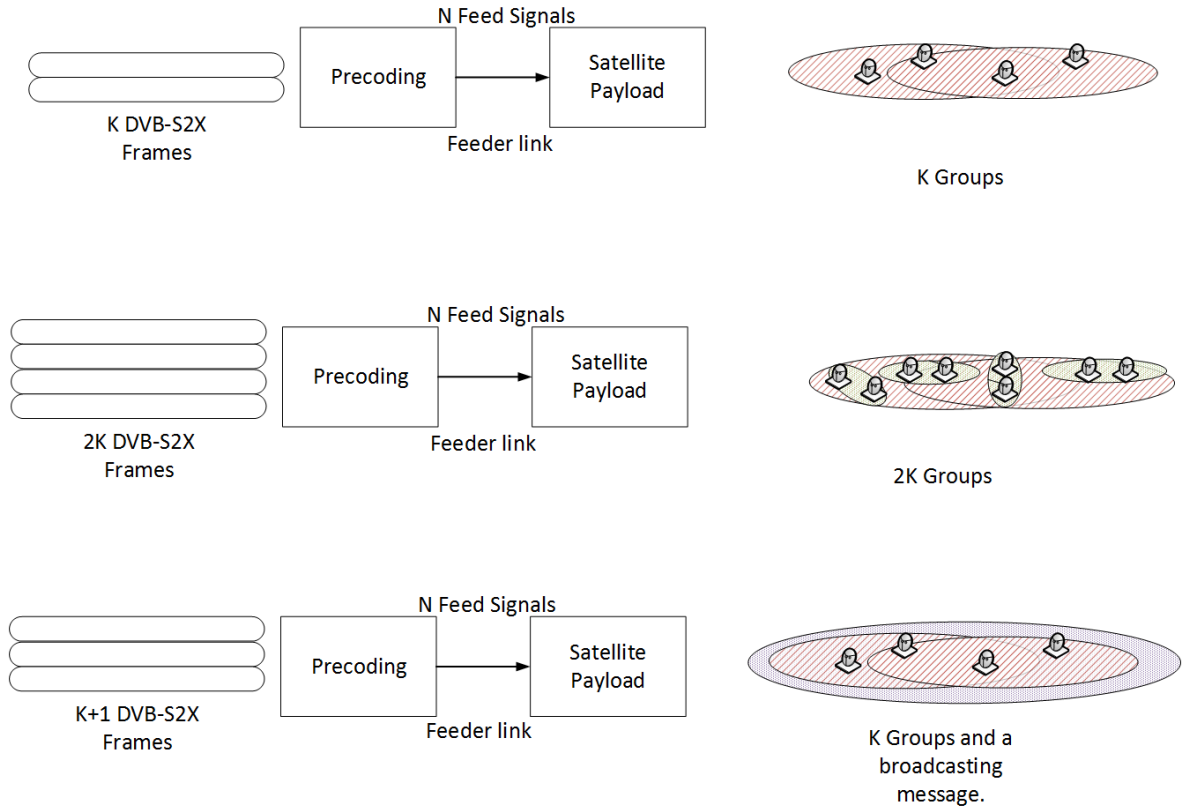

Figure 5: Up: Precoding in single layer transmission. Middle: Precoding in multi-layer transmission. Down: Precoding in joint broadcast and multicast transmission.

As a matter of fact, although the satellite payload can remain the same as the one used for multibeam precoding systems, it is clear that the feeder link, which goes from the gateway to the satellite, needs twice its capacity in order to attend the new system capacity for the multilayered transmission. Indeed, while, in a single layer, precoding transmission needs a total feeder link bandwidth of *BK* Hz (with *B* as the user bandwidth), the LDM transmission requires *2B.K* Hz. This restriction might involve the creation of twice the gateways or the utilization of higher frequency bands where more bandwidth is available. Similarly, the joint broadcast and multicast scheme requires *2B(K+1)* Hz of feeder link bandwidth.

Concerning user clustering in multicast precoding for multibeam satellite, [15] studies different alternatives in order to obtain high achievable rates. The user grouping is mainly based on the minimum Euclidean (minEuclidean) distance of the different user terminal channel vectors within the same beam. In particular, users with very low Euclidean distance between their channel vectors are served simultaneously. The underlying idea of this heuristic procedure is to facilitate the multicast precoding operation.

However, as we have commented before, in NOMA precoded systems the mentioned minEuclidean approach shall be revisited. Bearing in mind the different operations at the user terminal, it is unclear whenever minEuclidean will benefit them. Despite that the interbeam interference will be reduced by this approach, it remains an open issue whenever the intrabeam signal power levels, which rule the attainable rate of that beam, will benefit from this approach. As a matter of fact, and as we have commented, it is known that in cellular multilayer NOMA systems with

MUD or SUD receiver strategy, it is beneficial the power imbalance between different users. This power imbalance cannot be obtained by the legacy minEuclidean scheme as this scheme tends to gather users with similar channel gains and; thus, prospective equivalent power strength after precoding.

As an example, Fig. 6 depicts a system level evaluation of a multibeam satellite system with $N$ = 245 feed elements serving $K$ = 245 beams and a total number of users $2KN_U$ = 490 users. This is, we consider that $N_U$ = 1. We assume a maximum power per feed of 55 Watts and a power amplifier output back-off of 5 dB. The channel model is the one reported in [11], where the antenna pattern has been shared by the European Space Agency.

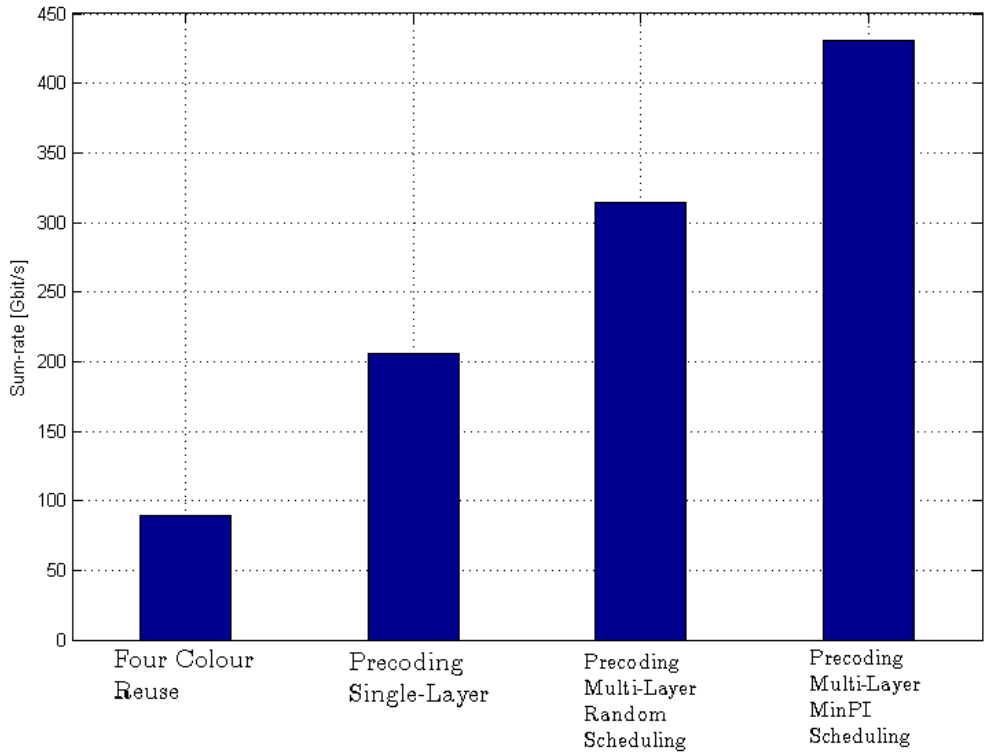

Figure 6. Sum rate evaluation for multibeam precoding transmission jointly with NOMA and different scheduling strategies.

Two benchmark cases have been included: the four-frequency scheme and the precoding with single layer transmission. Note that in both cases it has been considered a two time slot transmission in order to serve the 490 users with the considered 245 feed elements. The numerical results show that even with any type of scheduling, precoding with a multilayer transmission presents a substantial sum rate gain compared to the benchmark case. In case the mentioned minimum power imbalance scheduling is used, a 109 and 384 percent gain is obtained with respect to the single layer precoding case and the four-frequency scheme.

IV. **OPEN RESEARCH CHALLENGES AND CONCLUSIONS**

This article has reviewed the state-of-the-art in NOMA for satellite mulitbeam communications and has provided a taxonomy of the main representative scenarios with respect to the order of the frequency reuse. Remarkably, it is still under research which is the most convenient way to distribute the network resources (i.e. power, beam and bandwidth), in order to take advantage of the potential flexibility of V/HTS and increase the access network capacity with NOMA. We have noted that the framing of the multiuser data that is used in the satellite protocols is different from the one used in the terrestrial wireless standards. This is because current satellite communication systems need larger channel coding gains than terrestrial counterparts; each simultaneous frame transmission generally embeds more than one user terminal information, leading to the so-called multigroup multicast multibeam operation. This different framing has important consequences in the user scheduling and rate allocation. Therefore, these aspects motivate the need for different scheduling techniques as the one in [8], where the authors move from the work performed in [14] and propose a geographical scheduling algorithm in order to improve the fairness of a multicast precoder.

This article has shed some light on the role of NOMA in these flexible satellite systems; however, many aspects are yet to be defined and studied. There are pure physical layer aspects as: the impact of channel estimation errors at the receiver, the effect of timing offset between the signals transmitted by the different beams, low-complexity non-orthogonal coding/pre-coding and decoding strategies that leverage limited feedback or, a dynamic user scheduling and grouping strategies, to mention a few. These studies have to be cast with upper layers aspects, such as: non-uniform traffic distribution among beams, packet encapsulation and adaptive coding and modulation in NOMA, centralized or decentralized resource allocation, among others. The major advances have been carried out for centralized schemes. However, concerning decentralized ones, promising recent developments are within game theory as it was reported in [1] for the wireless cellular communications.

Finally, this article has focused on the forward link for V/HTS. However, there can also be interesting applications of NOMA in the reverse link, where multiple users have to access the satellite. In this respect, there are interesting and useful studies on random access with collision resolution for machine to machine satellite communications (i.e. enhance-spread spectrum aloha, E-SSA). In case of high SNR imbalance among users in the return link, grant-based (or dedicated access) schemes can also benefit from NOMA by allowing overlapping signal decoding at the receiving gateways. All these represent new research avenues.

## ACKNOWLEDGEMENTS

The authors thank SatnexIV Network of Excellence, from the European Space Agency for the fruitful discussions. This work has received funding from the Spanish Government under grant TEC2017-88373-R (TERESA) and from the Catalan Government (2017-SGR-1479).